\documentclass[11pt]{article}
\pdfoutput=1 
\usepackage{jinstpub} 
\usepackage{subfigure}

\title{Helical Six-Dimensional Muon Ionization Cooling Channel with Gas-Filled RF Cavities}

\author{Katsuya Yonehara}

\affiliation{Fermi National Accelerator Laboratory,\\Batavia, IL 60510, USA}

\emailAdd{yonehara@fnal.gov}

\abstract{}

\begin{document}
\maketitle
\flushbottom

\section{Introduction}
\label{sec:intro}

\begin{figure}[h]
\centering
 \includegraphics[width=0.8\textwidth,trim={2.4in 1.3in 2.8in 0.7in},clip]{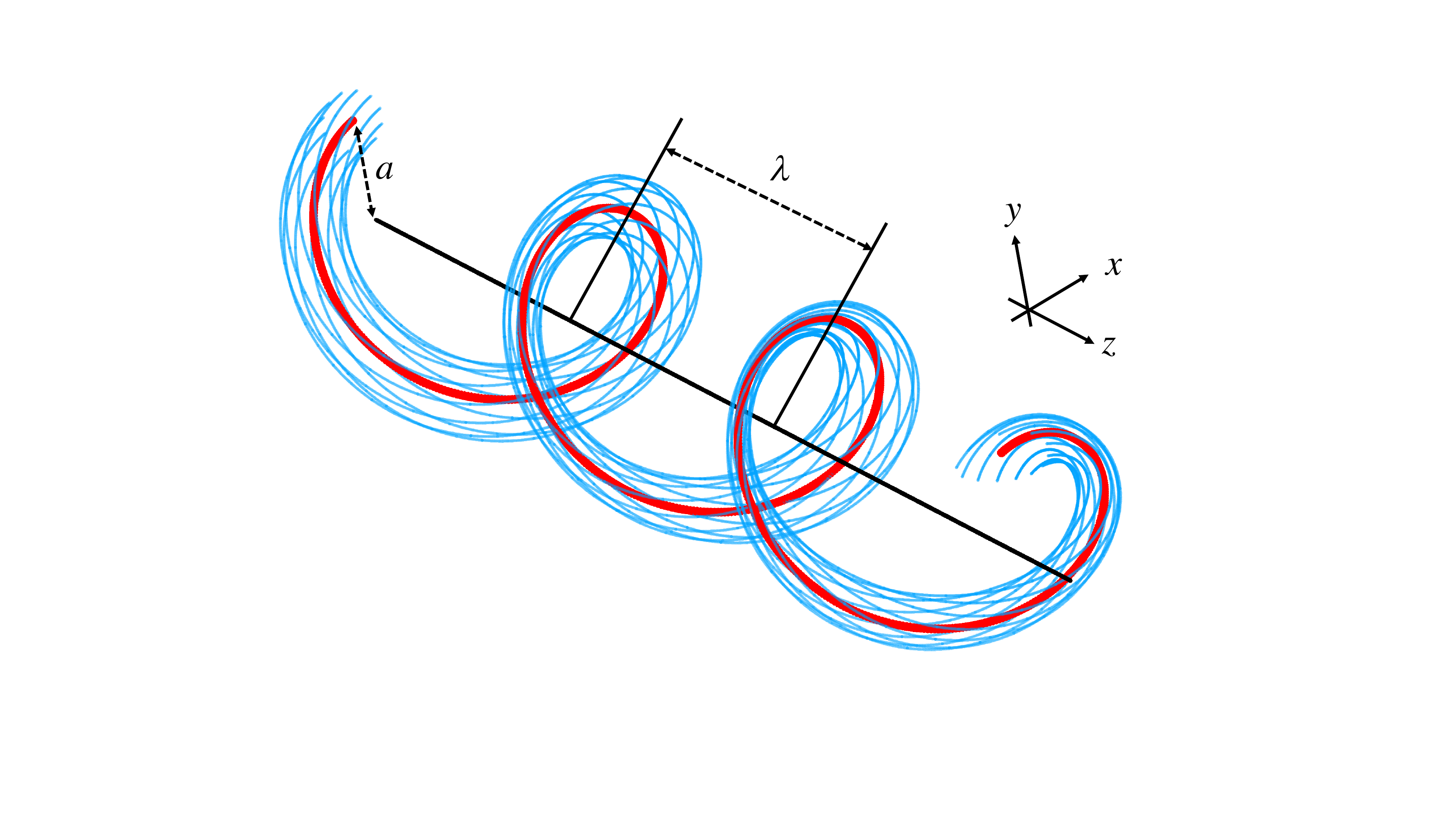}
 \caption{Typical beam path in a HCC.}
 \label{fig:orbit}
\end{figure}

A helical cooling channel (HCC) has been proposed~\cite{Slava05}, which consists of a helical magnet, filled with a homogeneous absorber that provides beam cooling by ionization energy loss.
The primary magnetic components in the HCC are the solenoid and counteracting helical dipole to define the reference trajectory and the helical dipole gradient that controls the dispersion and provides transverse stability. Figure~\ref{fig:orbit} shows the reference orbit (red line) and beam envelope (light blue line ensemble) of off-reference momentum particles in a channel comprised of HCC magnetic field components.
The reference particle moves on a helical trajectory around the Larmor center (black line), and 
the envelopes show coupled transverse and longitudinal oscillations around the reference orbit. 
An implementation of a helical dipole magnet is the ``Siberian Snake'' that has been used to 
manipulate the spin of polarized particles. 
The linear beam dynamics theory in the HCC has been validated using analytical~\cite{Slava05} and numerical methods. 
The concept of helical beam dynamics has been extended to apply for other beam elements, like phase space matching and isochronous bunch recombination channels. 

High-pressure hydrogen gas-filled RF (HPRF) cavities are another key element of the HCC. 
RF cavities are placed along the helical beam path in the HCC magnet. 
High-pressure hydrogen in the cavity acts as the homogeneous ionization absorber. 
Gaseous hydrogen is the best material because of its long radiation length and large energy loss rate, which together result in a low equilibrium emittance. 
The high-pressure gas serves a second purpose in reducing the probability of electric breakdown in the RF cavity and allows higher operating electric fields within the strong solenoidal magnetic field. 
In the case of a vacuum RF cavity, the probability of breakdown is amplified by the external magnetic field~\cite{Moretti, Palmer}. On the other hand, the dark current in the gas-filled RF cavity is insensitive to the external magnetic field since the dark current is suppressed immediately via Coulomb scattering in the gas (Paschen's law). The breakdown suppression model has been experimentally verified, and no RF degradation due to the external magnetic field has been observed ~\cite{Hanlet06}. 
High-pressure hydrogen gas provides another benefit for ionization cooling. The incident charged particle  beam ionizes the gas and generates a gas-plasma sheath in the cavity~\cite{freemire2016}, which
neutralizes the space charge of the beam by polarizing the plasma charge state. Consequently, the beam is focused by the self-induced toroid magnetic field and the beam is cooled by itself more than by conventional cooling optics. 

Since the HCC does not adhere to conventional beam optics, we'll note its specific prospects:
\begin{itemize}
    \item \textbf{Large momentum acceptance.}
    Since the optics of the HCC is continuous, there is no betatron tune resonance. 
    Thus, the HCC achieves a large momentum acceptance. 
    This tolerance permits lower RF gradients than those by other types of cooling channels. 
    \item \textbf{Cost-effective compact channel.}
    Since all ionization cooling processes, including RF acceleration and the emittance exchange, take place simultaneously in the HPRF cavity, the HCC can be the shortest design for a 6D ionization cooling channel. Note that a shorter channel simultaneously reduces costs and muon decay losses, resulting in increased performance per unit cost. 
    \item \textbf{Required novel beam element.}
    The HCC involves several novel technologies, of which the HPRF technology is one example. 
    Another challenging issue is to incorporate the HCC cavity system inside the compact helical magnet. Also, geometry constraints limit its cooling performance. 
    Understanding the limits and addressing them are a major part of the effort needed to design the HCC and to evaluate its performance via simulations. 
\end{itemize}
First, the concept of the HCC will be discussed to describe its unique features. 
Then, the current progress on design and studies of the HCC will be summarized. 

\section{Beam dynamics in HCC}
A complete linear beam dynamics theory in a helical field has been developed~\cite{Slava05}. 
An essential part of the HCC theory is introduced here. 
These formulas help us extrapolate the HCC theory to a practical conventional cooling system.
A stable particle orbit is always found as the reference orbit in the combined field of a solenoid ($B$) field with a helical dipole ($b$) component, 
\begin{eqnarray}
    p(a)&=&\frac{\sqrt{1+\kappa^2}}{k}\left( B-\frac{1+\kappa^2}{\kappa} b \right),
    \label{eq:equ}
\end{eqnarray}
where $a$ is the radius of the reference orbit from the Larmor center and $p$ is the reference momentum.
$\kappa$ is a helical pitch, which is fixed by the geometric constraint, $\kappa = 2\pi a/\lambda = ka$, where $\lambda$ is the helical period. 
$\kappa$ is also given by the ratio of the transverse and longitudinal momenta of the reference particle, $\kappa = p_\phi/p_z$. 
The dispersion factor, $(da/a)/(dp/p)$, is derived by differentiating Eq.~\eqref{eq:equ} with respect to $a$, 
\begin{eqnarray}
    \hat{D}^{-1} &=& \frac{\kappa^2 + (1-\kappa^2)q}{1+\kappa^2}+g, \label{usss} \\
    g &=&-\frac{(1+\kappa^2)^{3/2}}{pk^2}\frac{\partial b}{\partial a}, \label{eq:g}
\end{eqnarray}
where $\partial b/\partial a=b^\prime$ is the helical field gradient and $g$ is the field index. $q$ represents a coupling strength between horizontal and vertical motion in the helical coordinate system. $q$ is fixed from the dispersion, which will be given in later section. 

The transverse beta tunes are given as eigenvalues of the second-order equation of motion, 
\begin{eqnarray}
    Q^2_{\pm}&=&R\pm \sqrt{R^2-G}, \label{eq:beta} \\
    R&=&\frac{1}{2}\left(1+\frac{q^2}{1+\kappa^2} \right), \\
    G&=&\left( \frac{2q+\kappa^2}{1+\kappa^2}-\hat{D}^{-1} \right) \hat{D}^{-1}.
\end{eqnarray}
The transverse beta functions are 
\begin{equation}
    \beta_{\pm} =\frac{1}{kQ_{\pm}}=\frac{\lambda}{2\pi Q_{\pm}}.
    \label{eq:betaT}
\end{equation}
The beam stability condition in the transverse phase space, $0 < G < R^2$ is derived from Eq.~\eqref{eq:beta}. 
On the other hand, the longitudinal beta function is 
\begin{equation}
    \beta_L = \frac{1}{\omega Q_L} = \sqrt{\frac{m_\mu c}{\eta \omega e V^{\prime}}} 
    \frac{1+\sin (\phi_s)}{1-\sin(\phi_s)},
    \label{eq:betaL}
\end{equation}
where $V^{\prime}$ is the peak RF gradient, $m_{\mu}$ is the muon mass, and $\phi_s$ is the synchrotron phase. $\eta$ is the momentum slip factor, 
\begin{equation}
    \eta =\frac{d}{d\gamma} \frac{\sqrt{1+\kappa^2}}{\beta} =\frac{\sqrt{1+\kappa^2}}{\gamma\beta^3} 
    \left( \frac{\kappa^2}{1+\kappa^2}\hat{D} -\frac{1}{\gamma^2} \right).
\end{equation}
The general solution of particle motion in the helical coordinate system is the sum of those three eigenfunctions.

\section{Cooling performance of HCC}
\begin{figure}[h]
 \includegraphics[width=\textwidth,trim={0 2in 0 1.2in},clip]{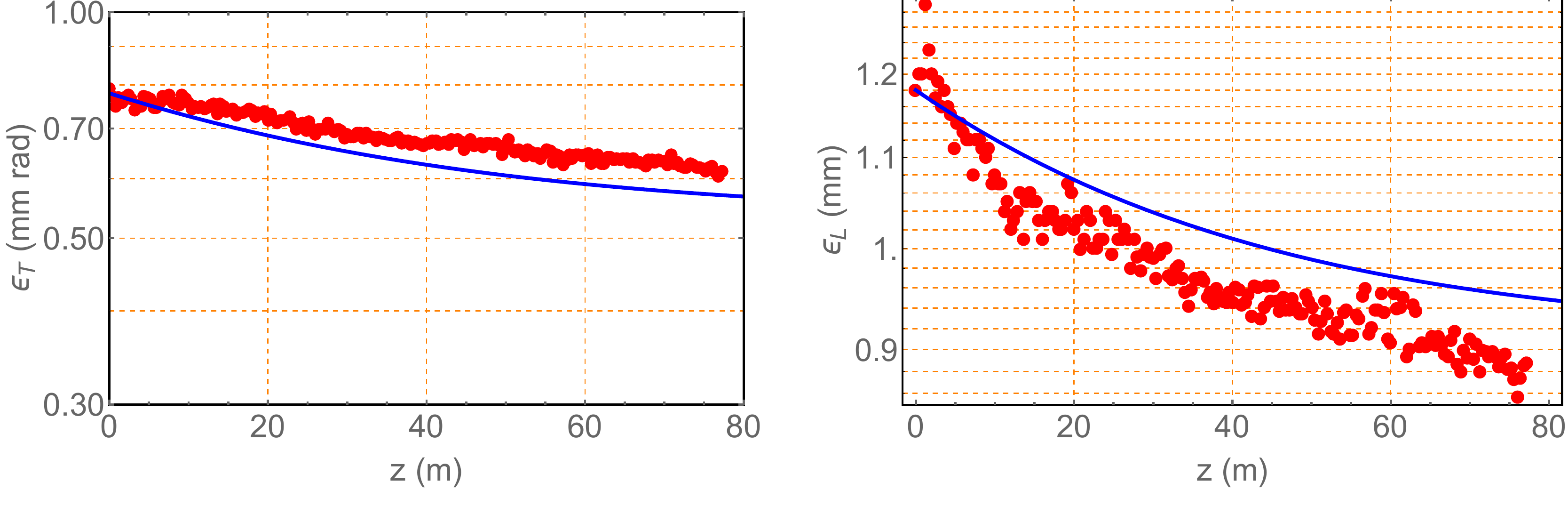}
 \caption{Red dots are results from a simulation for the last segment of the HCC (segment 4) while the blue line is the prediction from theory.
}
 \label{fig:theory_vs_simulation}
\end{figure}

The helical magnetic component is fixed once a reference momentum $p$, helical pitch $\kappa$, helical period $\lambda$, and dispersion factor $\hat{D}$ are chosen. 
Table~\ref{tab:field} shows the summary of cooling simulation results in the HCC using G4Beamline~\cite{Tom} on NERSC~\cite{nersc}. 
\begin{table}[h]
\caption{Field parameters and normalized beam emittances in the end-to-end cooling simulation. $\epsilon_{tr}$ is the transmission efficiency for each HCC segment. 
The majority of loss is due to muon decay. $B_z = B-\kappa b$, where $B$ is provided by a pure solenoidal magnet and $\kappa b$ is generated by the helical conductor as given by Eq.~\eqref{eq:equ}.  
From optimization of the HCC, $\kappa$ is 1, the reference momentum is 200 MeV/c, and the dispersion factor is 1.81 in this simulation. $q$ is 0.87 from the dispersion factor. Peak RF gradient in the gas-filled RF cavities is 20 MV/m. Hydrogen gas pressure is 160 atm at room temperature. 30 $\mu$m-thick Be RF windows are located at both ends of the RF cavity. The number of RF cavities per helical period is 10. 
}
\label{tab:field}
\centering
\begin{tabular}{|llllllllll|}
\hline
Seg. &\  $\lambda$ &\ $L$ &\  $\nu$ &\ $B_z$ &\ $b$ &\ $b^\prime$ &\ $\varepsilon_{T,eq}$ &\ $\varepsilon_{L,eq}$ &\  $\varepsilon_{tr}$ \\
\hline
unit &\ m &\  m &\ MHz &\ T &\ T &\ T/m &\ mm rad &\ mm &\ \\
\hline
1 &\ 1.0 &\  50 &\ 325 &\ 4.41 &\ 1.32 &\ -0.32 &\  3.44 &\ 6.82 &\ 0.94 \\
2 &\ 0.8 &\  70.4 &\ 325 &\ 5.52 &\ 1.65 &\ -0.50 &\  1.62 &\  2.41 &\ 0.90 \\
3 &\ 0.5 &\  120 &\ 650 &\ 8.83 &\ 2.63 &\ -1.28 &\  0.79 &\  1.18 &\ 0.81 \\
4 &\ 0.4 &\  77.2 &\ 650 &\ 11.04 &\ 3.29 &\ -2.01 &\  0.61 &\  0.89 &\ 0.85 \\
\hline
 &\ &\ 317.6 &\ &\ &\ &\ &\ &\ &\ 0.58 \\
 \hline
\end{tabular}
\end{table}
The exact HCC theory predicts the emittance evolution as shown in Fig.~\ref{fig:theory_vs_simulation}. 

The cooling formula in the HCC theory is complicated, since it has a tangled term among three eigenvectors. 
Fortunately, the tangled term contributes $\sim$20\% in the exact formula. 
Here, we attempt to extrapolate HCC theory to a conventional cooling scheme by ignoring the tangled term. 
According to Neuffer's cooling scheme~\cite{Neuffer13}, a degree of emittance exchange, namely the cooling partition, can be represented for the homogeneous ionization absorber filled HCC as 
\begin{equation}
\delta g_{L} = \frac{\kappa^2}{1+\kappa^2}\hat{D}.
\label{eq:part}
\end{equation}
The conventional emittance evolution formulae are, 
\begin{eqnarray}
    \frac{d\varepsilon_T}{ds}&=&-\frac{g_T}{\beta^2 E}\frac{dE}{ds} \varepsilon_T 
    +\frac{\beta\gamma}{2}\beta_T\frac{d \langle \theta^2_{rms} \rangle}{ds}, \label{eq:emitt} \\
    \frac{d\varepsilon_L}{ds}&=&-\frac{g_L}{\beta^2 E}\frac{dE}{ds} \varepsilon_L 
    +\frac{\beta\gamma}{2}\beta_L\frac{d \langle \left( \delta p/p \right)^2_{rms}\rangle}{ds}, \label{eq:emitl}
\end{eqnarray}
where 
\begin{eqnarray}
    g_L &\rightarrow& g_{L,0}+\delta g_L, \label{eq:gl}\\
    g_T &\rightarrow& 1-\frac{\delta g_L}{2}, \label{eq:gt}\\
    g_{L,0}&=&-\frac{2}{\gamma^2}+2 \frac{\left( 1- \beta^2/\gamma^2 \right)}{\ln \left( 2 m_e c^2 \beta^2 \gamma^2/I (Z) \right)}. \label{eq:gl0}
\end{eqnarray}
By substituting Eqs.~\eqref{eq:betaT}, \eqref{eq:betaL}, \eqref{eq:part} into Eqs.~\eqref{eq:emitt}, \eqref{eq:emitl}, \eqref{eq:gl}, \eqref{eq:gt}, \eqref{eq:gl0}, all cooling parameters in the HCC
are represented using the conventional cooling notation. 

\section{Features of HCC} 
The following list compares the features of the HCC to other cooling channels. 
\begin{itemize}
    \item \textbf{Transverse cooling.}
    The dominant focusing field in the HCC is a continuous solenoidal field. It is known that the transverse beta function in a continuous solenoid channel is twice as long as that in an alternate solenoid one. 
    Indeed, the beta function in the HCC is two times longer than that in the rectilinear channel at the same solenoid field strength. Consequently, the equilibrium transverse emittance in the HCC is twice as large as that in the rectilinear channel. 
    \item \textbf{Longitudinal cooling.}
    The longitudinal beta function in the HCC can be made shorter than that in the rectilinear channel by increasing the momentum slip factor. Indeed, the simulation shows that the equilibrium longitudinal emittance in the HCC is smaller than in the rectilinear channel. 
    \item \textbf{Momentum slip factor.}
    Momentum slip factor in the HCC is positive. The upstream and downstream beam optics from the HCC are typically a solenoid base channel, which has a negative momentum slip factor. 
    Thus, a longitudinal phase space matching channel is required. 
    The transmission efficiency in the present matching channel for the HCC is $\sim$80\%~\cite{Yoshikawa13}. On the other hand, the momentum slip factor in the HCC can be fixed o zero by changing the helical coil configuration. This suggests that the helical channel becomes an isochronous channel. This can also be applied to a bunch recombination channel.  
    The channel is short, effective and practical~\cite{Cary14, Sy15}. 
\end{itemize}

\begin{figure}[h]
\centering
 \includegraphics[width=8cm]{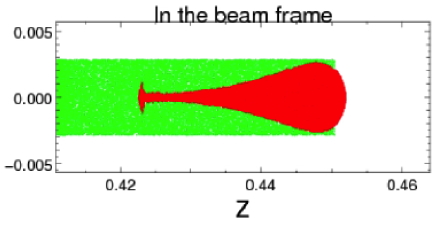}
 \caption{Simulation of beam dynamics in the gas-filled RF cavity. 
 Red points are the incident beam, and the green ones are the ionization electrons.
 The front of the beam generates ion pairs, and the back end is focused by the plasma lens effect. 
   }
  \label{fig:lens}
\end{figure}

Interestingly, while the space charge effect influences the beam dynamics, the collective force focuses the beam in the HCC. The longitudinal space charge force naturally focuses muons in the longitudinal direction because the HCC has a positive momentum slip factor. On the other hand, the beam-induced gas plasma in the RF cavity will neutralize the space charge. The plasma is polarized by the space charge field and forms a plasma sheath along the beam path. 
As a result, the self-induced toroidal magnetic field focuses the beam~\cite{Kwang14}. 
Both focusing forces are stronger in a denser beam. 
This represents a new cooling mechanism and generates extra cooling. 
Figure~\ref{fig:lens} shows the transverse focusing of the beam in the gas-filled RF cavity that is produced by the space charge neutralization effect~\cite{Ellison}. 
The front-end beam generates plasma, which neutralizes the space charge in the rest of beam, and the self-induced field pinches the back-end beam. 
Since the simulated channel is a simple solenoid field, the longitudinal focusing is not involved. 
Note that only linear plasma dynamics are used in this simulation. 
Non-linear dynamics should be included in further studies. 

\section{Design concept}
\begin{figure}[h]
\centering
 \includegraphics[width=0.6\textwidth]{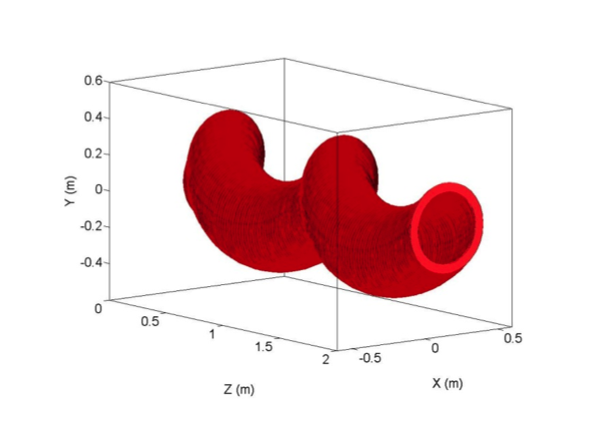}
 \caption{Schematic drawing of the helical solenoid magnet. 
    }
  \label{fig:hscoil}
\end{figure}

Once the dispersion is determined, all HCC field components on the reference orbit are fixed,
\begin{eqnarray}
    B_{z} &=& pk \frac{1+q(\hat{D})}{e \sqrt{1+\kappa^2}}, \label{eq:bz} \\
    b &=& \kappa B_z \frac{1-1/(1+q(\hat{D}))}{1+\kappa^2}, \label{eq:b} \\
    g &=& \hat{D}^{-1}-\frac{\kappa^2+(1-\kappa^2)q(\hat{D})}{1+\kappa^2}, \label{eq:g} \\
    \frac{\partial b}{\partial a} &=& b^\prime = \frac{-e g p k^2}{(1+\kappa^2)^{3/2}}. \label{eq:bp}
\end{eqnarray}
Initially, a helical conductor (like the Siberian snake) was considered. 
However, we noticed that the beam occupies only a quarter of the magnetic field in the helical dipole conductor. 
A helical solenoid coil has been proposed to generate the required helical field in the beam area. 
Figure~\ref{fig:hscoil} shows a schematic drawing of the helical solenoid magnet. 
The helical solenoid coils are aligned along the helical beam path. 

\begin{figure}[h]
\centering
    \subfigure[Helical dipole component, $b$.]{
    \includegraphics[width=.48\textwidth]{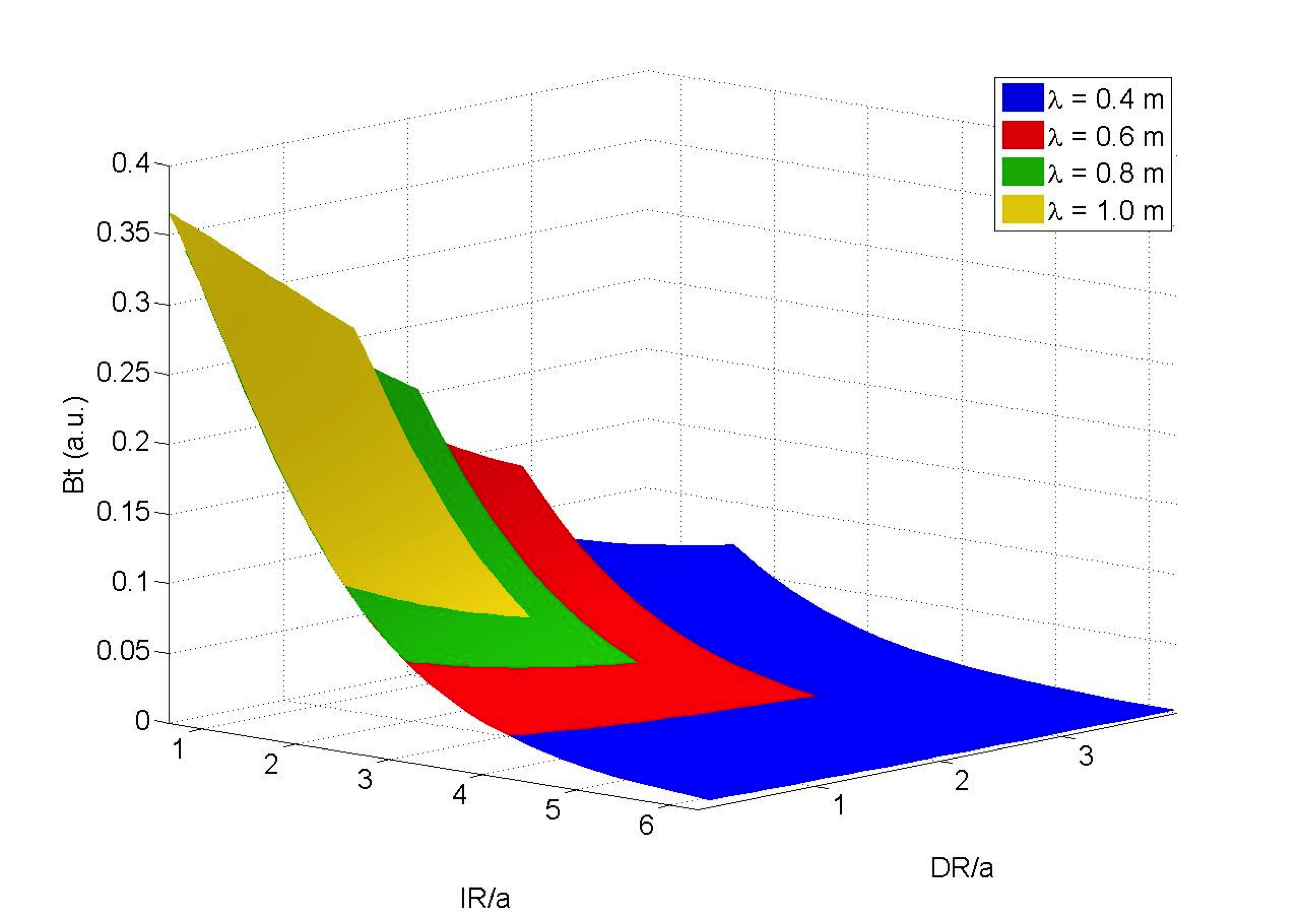}
    \label{fig:subfig1}
    }
    \subfigure[Helical field gradient, $b^\prime$.]{
    \includegraphics[width=.48\textwidth]{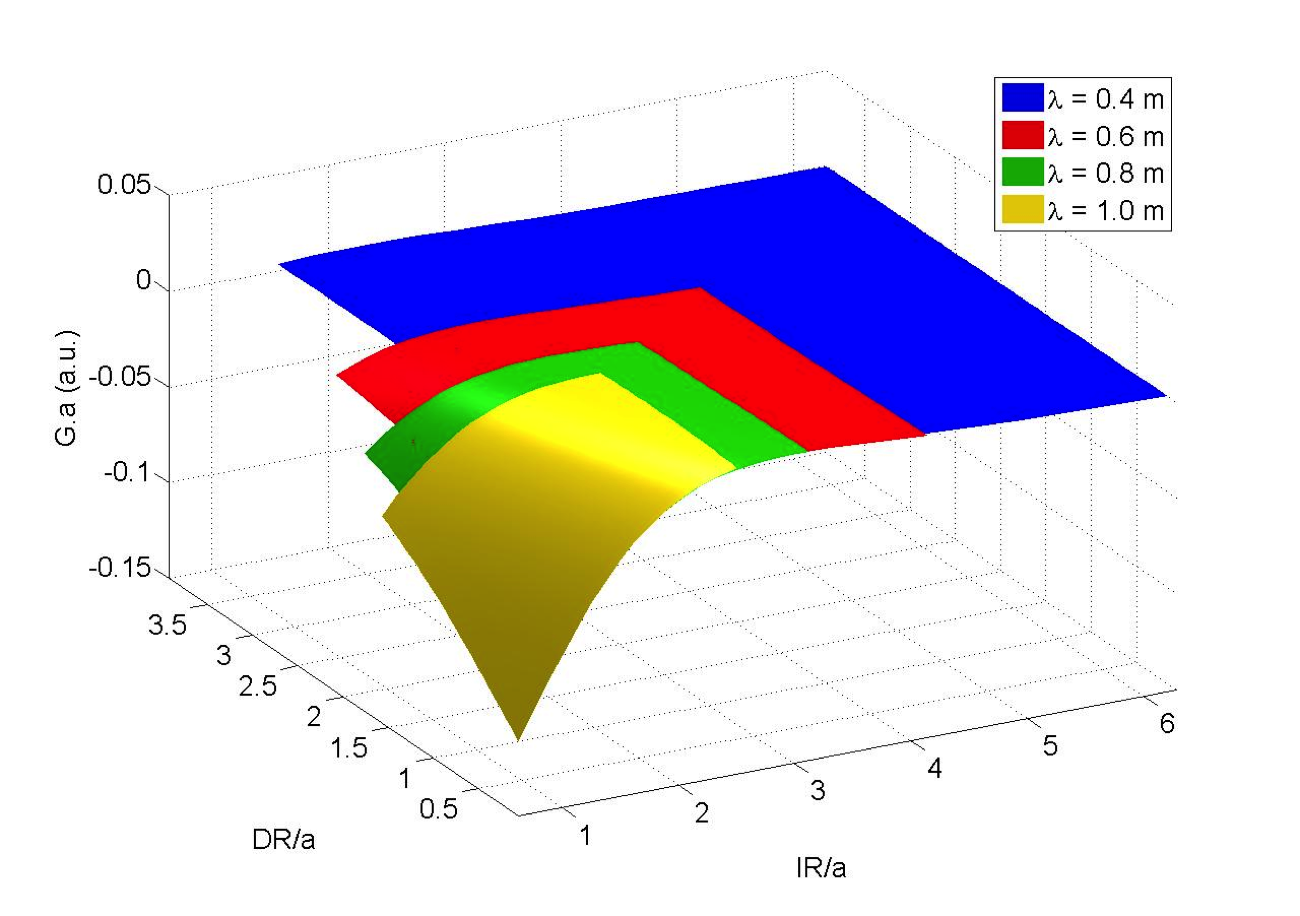}
    \label{fig:subfig2}
    }
\caption{$b$ and $b^\prime$ as a function of inner radius ($ID$) and coil radial thickness ($dR$) of the helical solenoid coil 
    normalized by the radius of the reference orbit ($a$). 
    A surface plot shows that the helical field component is independent of $\lambda$. 
    }
\label{fig:hsfield}
\end{figure}

The helical dipole component and the helical field gradient are generated by the stray field of the adjacent coils. Therefore, $b$ and $b^\prime$ are dependent on the geometry of the coils~\cite{Mau14}. 
Figure~\ref{fig:hsfield} shows the geometry constraints of $b$ and $b^\prime$, i.e. an inner coil diameter ($ID$), a radial thickness ($dR$), and the center of the coil ($a$). 
Several helical solenoid coils were made, and the concept was experimentally verified. 
The analysis suggests that it may be feasible to make the segment 4 in Table~\ref{tab:field}. 
Specific helical field components can be enhanced by using an elliptic-shaped coil, or a tilted coil is another option to generate the proper helical field~\cite{Kahn14}. 

\begin{figure}[h]
\centering
 \includegraphics[width=0.8\textwidth]{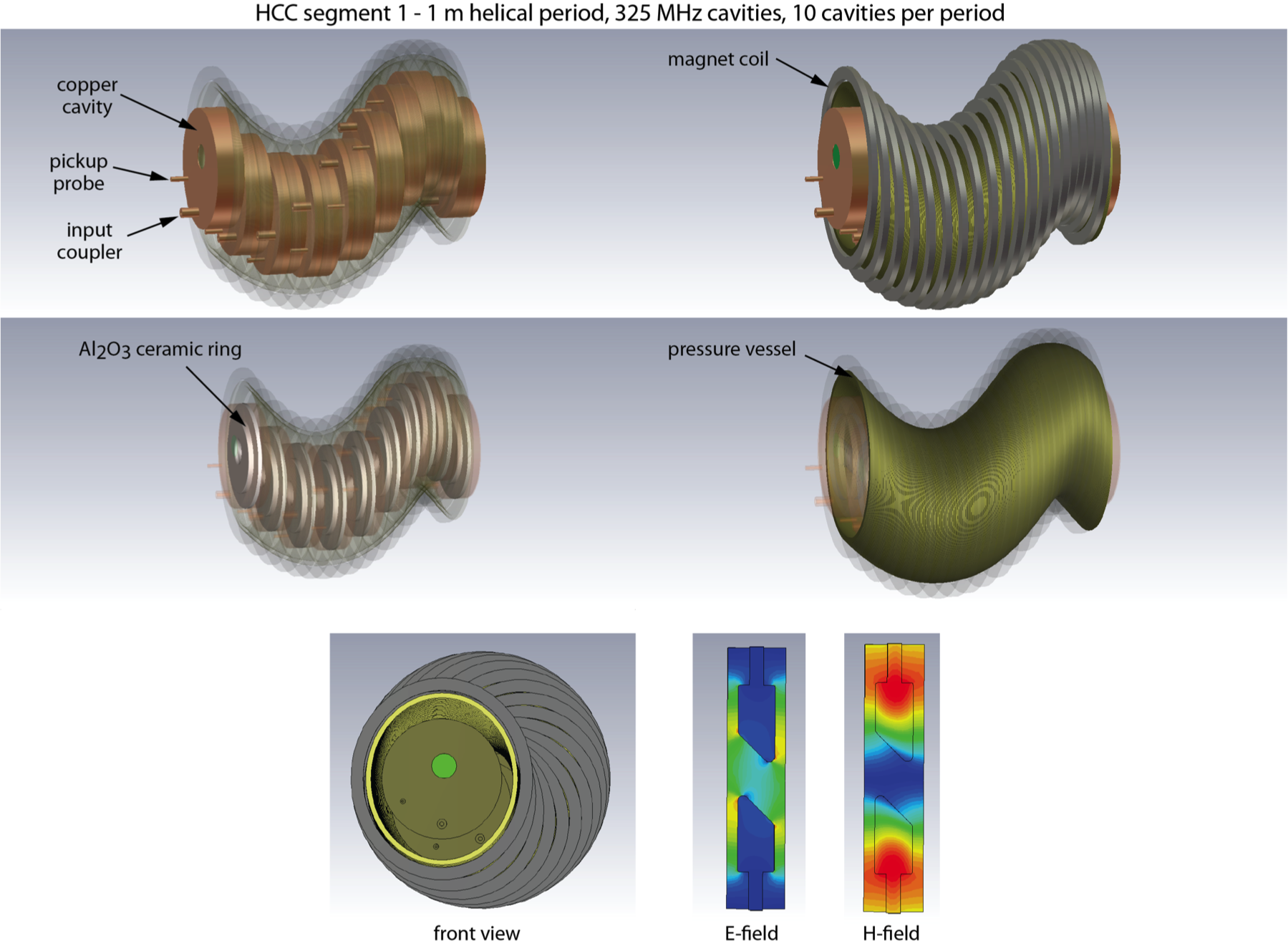}
 \caption{Schematic drawing of the segment HCC. 
    }
  \label{fig:integ}
\end{figure}

The HCC is designed to integrate the helical RF cavities in the magnet and to incorporate the other infrastructures. Figure~\ref{fig:integ} shows the design concept. 
To shrink the RF cavity, a dielectric insert cavity is proposed~\cite{Lisa13,Ben16}. 
30 $\mu$m-thick Beryllium is used as a beam entrance window. 
A thinner window is better suited to reduce the multiple scattering in the window~\cite{Yone14}. 
The minimum thickness of the window is limited to avoid detuning of the cavity due to Lorentz force and thermal expansion. 
Hydrogen gas is useful to remove heat from the window. 
Analytical investigations suggests that the RF window should be pre-stressed and curved-shaped~\cite{Alvin}. 
A demonstration test is needed. 

\section{Summary}
The concept of the helical cooling channel has been validated in analytical and numerical simulations. 
Several unique features in the HCC are reviewed, e.g. its large momentum acceptance, tunable momentum slip factor, response to the space charge effect, etc. In particular, the plasma focusing mechanism can be used for extra cooling of the muon beam. This technique is only applicable to muon beams.

%
%
%


\end{document}